\documentclass[twocolumn]{article}
\usepackage{graphicx}
\usepackage{amsmath}
\usepackage{titlesec} 
\usepackage{lipsum} 
\usepackage{geometry} 
\usepackage{float}
\usepackage{pdfpages}
\usepackage{dblfloatfix}
\usepackage{caption}
\usepackage{placeins}
\usepackage[numbers]{natbib}

\title{ONCOPILOT: A Promptable CT Foundation Model For Solid Tumor Evaluation}

\date{}

\begin{document}

\onecolumn
\maketitle

\begin{center}
\textbf{Direct contribution}: \\
Léo Machado (MD, PhD)\textsuperscript{1,2}, Hélène Philippe (MSc)\textsuperscript{1,2,3}, Elodie Ferreres (MSc)\textsuperscript{1}, Julien Khlaut (MSc)\textsuperscript{1,4}, Julie Dupuis\textsuperscript{1}, Korentin Le Floch (MD)\textsuperscript{1,4}, Denis Habip Gatenyo (MD)\textsuperscript{5} \\
[0.5em] 
\textbf{Senior contribution}: \\
Pascal Roux (MD)\textsuperscript{6}, Jules Grégory (MD, PhD)\textsuperscript{2,3}, Maxime Ronot (MD, PhD)\textsuperscript{2,3}, Corentin Dancette (PhD)\textsuperscript{1}, Tom Boeken (MD, PhD)\textsuperscript{4}, Daniel Tordjman (MSc)\textsuperscript{1}, Pierre Manceron (MSc)\textsuperscript{1}, Paul Hérent (MD, MSc)\textsuperscript{1,6} \\[0.5em]

\textbf{Affiliations:} \\
{\small
\textsuperscript{1}Raidium, Paris Biotech Santé, Paris, France \\
\textsuperscript{2}AP-HP. Nord, Department of Radiology, FHU MOSAIC, Beaujon Hospital, Clichy, France \\
\textsuperscript{3}Université Paris Cité, Paris, France \\
\textsuperscript{4}Université Paris Cité, AP-HP, Hôpital Européen Georges Pompidou, Department of Vascular and Oncological Interventional Radiology, HEKA INRIA, INSERM PARCC U 970, Paris, France \\
\textsuperscript{5}Department of Radiology, Hôpital Cochin, AP-HP, Paris, France \\
\textsuperscript{6}Centre d’Imagerie du Nord, Saint-Denis, France
}

\vspace{1em} 

\end{center}

\vspace{2em} 
\begin{center}
\noindent\rule{0.8\textwidth}{0.4pt} 
\end{center}

\vspace{2em}

\begin{center}
{\Large \textbf{Abstract}} 
\vspace{0.5em} 
\end{center}

Carcinogenesis is a proteiform phenomenon that can lead to metastatic spread, with tumors emerging in various locations and displaying complex, diverse shapes. As a crucial focus at the intersection of research and clinical practice, it demands precise and flexible assessment. However, current biomarkers, such as RECIST 1.1's long and short axis measurements, fall short of capturing this complexity, offering an only approximate estimate of tumor burden and an overly simplistic representation of a far more intricate process. 

Additionally, existing supervised AI models face challenges in adequately addressing the variability in tumor presentations, which limits their clinical utility. These limitations arise from the scarcity of annotations and the models' focus on narrowly defined tasks.

To address these challenges, we developed \textbf{ONCOPILOT}, an interactive radiological foundation model trained on approximately 7,500 CT scans covering the whole body, from both normal anatomy and a wide range of oncological cases. ONCOPILOT performs 3D tumor segmentation using visual prompts like point-click and bounding boxes, outperforming state-of-the-art models (e.g., nnUnet-based) and achieving radiologist-level accuracy in RECIST 1.1 measurements. The key advantage of this foundation model is its ability to surpass state-of-the-art performance while keeping the radiologist in the loop, a capability that previous models could not achieve. When radiologists interactively refine the segmentations, accuracy improves even further. ONCOPILOT also accelerates measurement processes and reduces inter-reader variability. Moreover, it facilitates volumetric analysis, unlocking new biomarkers for deeper insights.

This AI assistant is expected to enhance the precision of RECIST 1.1 measurements, unlock the potential of volumetric biomarkers, and improve patient stratification and clinical care, while seamlessly integrating into the radiological workflow.

\newpage
\twocolumn

\section{Introduction}
The wide variability in tumor appearance and location makes precise monitoring of oncological disease a critical challenge for both clinical care and research. Effective evaluation of oncological disease is essential for accurately assessing tumor aggressiveness, predicting prognosis, and guiding treatment decisions.

The Response Evaluation Criteria in Solid Tumors (RECIST v1.1) has long been regarded as the gold standard for radiologically assessing solid tumors over time \cite{eisenhauer2009}, allowing for patient stratification based on disease response or progression. However, this method has significant limitations:  the low information yield from linear long axis measurement in comparison to total tumor burden \cite{Meignan2016, Chung2014}, the arbitrary and non-reproducible selection of target lesions, which can result in the misclassification of disease status \cite{Kuhl2019} and significant inaccuracies in measuring the long axis, with inter-reader variability exceeding 20\% \cite{Yoon2016}, further contributing to classification errors. 

Traditionally, the long and short axes of the tumor are used as widely accepted proxies for estimating tumor size on CT scans. However, in the era of quantitative imaging, these linear measurements are increasingly considered inadequate as the field shifts toward more informative quantitative markers, such as volumetry \cite{Planz2019} and shape assessments, including tumor eccentricity and irregularity \cite{Wang2018}. Volumetric analysis, more sensitive to change than diameter due to its proportionality to the cube of the radius, is proving advantageous in detecting tumor burden changes, especially for tumors with irregular shapes, where linear measurements fail to capture their complexity \cite{Hayes2016}. Recently, novel radiomics biomarkers derived from volumetric analysis have shown significant promise in oncological evaluation, notably in colon and lung cancers \cite{Dercle2022, Dercle2023}.

Despite its promise, volumetric measurement is time-consuming \cite{Zimmermann2021} and impractical to perform manually. While efforts have been made to automate the volumetric delineation of oncological lesions, from early models relying on manual feature extraction to approaches using deep learning with convolutional neural networks \cite{tandon2024}, these models remain limited. Most are organ-specific, and the solutions currently employed in clinical practice are effective primarily in straightforward cases, such as lung nodules, but struggle with the diverse appearances of metastatic lesions. Furthermore, these methods often lack interactivity and adaptability to varying inputs, which restricts their integration into the radiological workflow.

The emergence of foundation models, a paradigm shift in deep learning, could alleviate these issues. Powered by transformer architecture and self-attention mechanisms \cite{Vaswani2017}, foundation models, when trained on extensive datasets, can significantly outperform traditional deep-learning systems \cite{Bommasani2021}. Their key strengths lie in transfer learning and zero-shot classification, enabling them to handle tasks not encountered during initial training—capabilities that traditional deep-learning models lack. 

Recently, these models have been positioned as the future of medical imaging, offering potential solutions to critical challenges such as poor generalization and the need for large quantities of labeled training data \cite{Azad2023}. Notably, they can generate reliable segmentation masks using text or visual prompts (e.g., actions taken on the images by the user), such as bounding boxes or point-click inputs on regions of interest \cite{Kirillov2023}. The ability to dynamically refine segmentation masks and generate varying outcomes from different visual prompts is a crucial step toward explainable AI, addressing the opaque nature of traditional models, and making it usable for radiologists.

In response to these advancements, we developed ONCOPILOT, an interactive foundation model trained using publicly available CT scans of normal anatomy and more than 7,500 tumors from diverse organs. Our model aims to deliver precise and reproducible RECIST measurements and to facilitate the volumetric analysis of oncologic lesions within an interactive viewer. We evaluated ONCOPILOT against a panel of radiologists and investigated the potential for integrating this AI assistant into the radiologist's workflow.

\section{Materials and Methods}

\subsection{Foundation Model}

ONCOPILOT is a foundation model adapted from SAM \cite{Kirillov2023}, specifically aiming to segment biomedical images. Such an approach has been concomitantly adopted in the literature in the form of MedSAM \cite{Ma2024}, SegVol \cite{Du2023}, and SAM-Med3D \cite{Wang2023}. This model is trained to perform image segmentation tasks. It processes 2D images and prompts, such as a bounding box, a point, or a mask. The model aims to generate a 3D prediction of the volume of a specific anatomical structure based on the input image and visual prompt. From the initial slice 2D segmentation masks are propagated sequentially across the z-axis until they encounter the boundaries of the object. Alternatively, the propagation can be halted based on predefined stopping criteria once specific conditions are met. The result of this propagation is a 3D segmentation mask. 

The model is initialized with the released weights of the SAM model \cite{Kirillov2023} and trained in a supervised manner with an initial general pre-training on normal anatomy and oncological lesions followed by a specialization on oncological lesions with a focused fine-tuning on tumors only. The initial training took 40 hours on 32 V100 Nvidia GPUs (totaling 1280 GPU hours) with a learning rate of $10^{-5}$. The fine-tuning took 10 hours on a Nvidia 4090 GPU. 

\begin{figure*}[htbp]
    \makebox[\textwidth][c]{\includegraphics[width=0.7\textwidth]{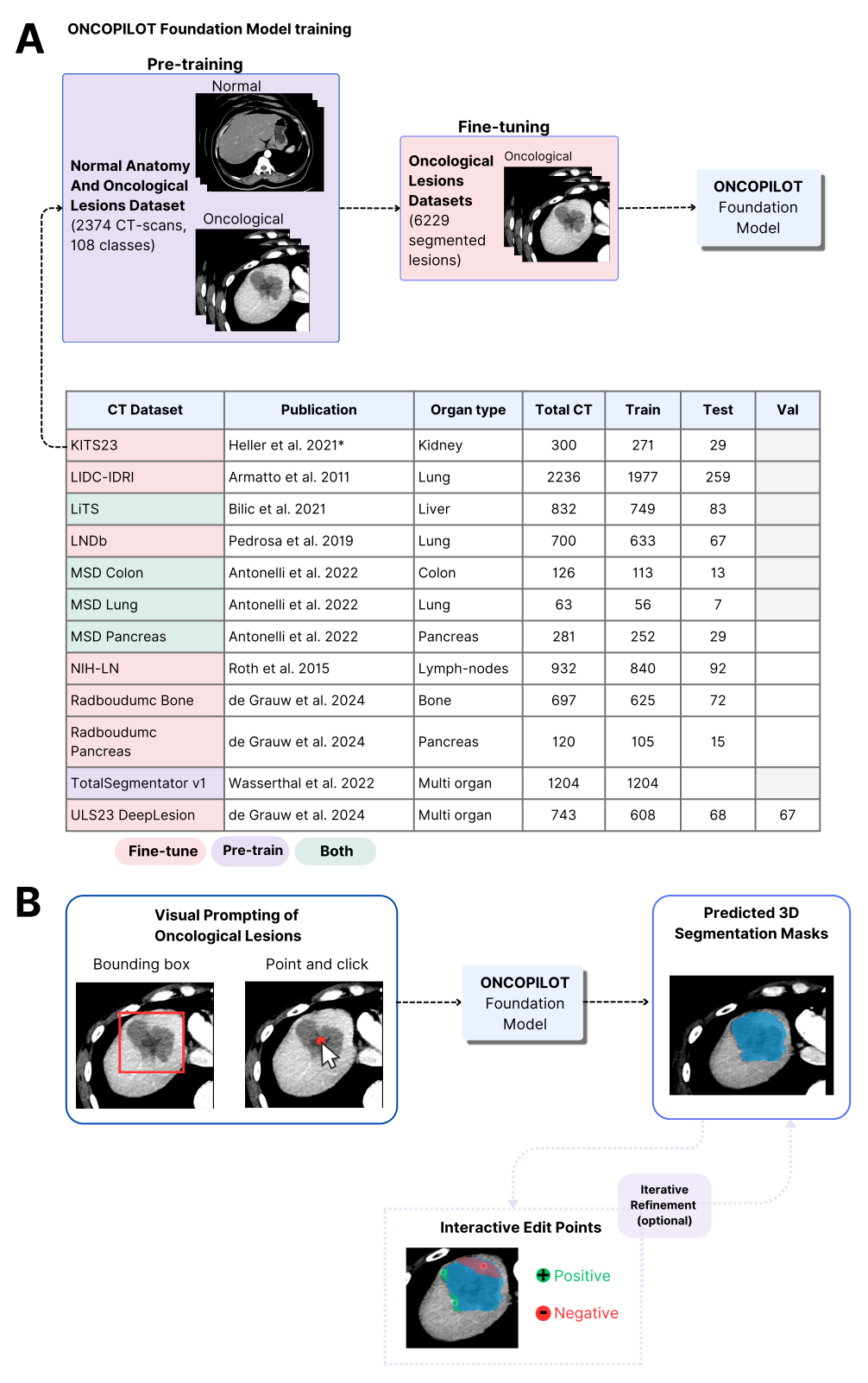}}
    \captionsetup{width=1\textwidth, labelfont=bf}
    \caption{\textbf{ONCOPILOT Foundation Model Training and Evaluation} (A) Overview of the datasets used for training the ONCOPILOT segmentation model, including the distribution across train, test, and validation sets. (B) Diagram illustrating the ONCOPILOT segmentation model's workflow. The model accepts visual prompts (either point-clicks or bounding boxes) of 3D tumor volumes and outputs corresponding 3D segmentation masks. Optional editing allows for real or simulated radiologist interaction, where positive and negative edit-points can be set manually in a viewer environment or automatically during evaluation.}
    \label{fig:fig1}
\end{figure*}

\subsection{Evaluation criteria of ONCOPILOT’s performances}

To assess the effectiveness of our model, three parameters are taken into account:

\textbf{Segmentation performance}: Measured using the DICE score, a standard metric for evaluating segmentation quality. The performance of ONCOPILOT is compared to that of a state-of-the-art model, see section \textit{Baseline}.

\textbf{Long axis measurement performance}: Evaluated using the absolute error and inter-operator variability as metrics. For more details, see section \textit{RECIST measurement and ONCOPILOT evaluation against radiologists}.

\textbf{Radiologist’s time efficiency and precision}: The average time a radiologist takes to complete one measurement and the inter-reader variability serve as metrics for evaluating the integration of the model into a radiologist's workflow. For further information, see section \textit{ONCOPILOT integration into radiologist’s workflow}.

\subsection{Baseline}

To compare the performance of our foundation model to state-of-the-art segmentation models we used the model provided by the ULS23 oncological lesion segmentation challenge as a baseline \cite{Grauw2024}. We used the result of their full model (nnUnet-ResEnc+SS) on the 10\% held-out test set originating from their fully-labeled dataset. This model is based on the nnUnet architecture \cite{Isensee2021} and has been trained and fine-tuned on a dataset of 38,693 lesions including fully-labeled and partially-labeled tumor masks.

\subsection{Datasets}
ONCOPILOT was pre-trained on publicly available datasets containing medical images and segmentation masks for general anatomy and oncological lesions:
\begin{itemize}
    \item 1204 CT scans from TotalSegmentator v1 \cite{Wasserthal2022}, with 104 labeled anatomical structures (27 organs, 59 bones, 10 muscles, 8 vessels).
    \item 743 diverse tumors from the DeepLesion dataset \cite{Yan2018}, curated and segmented for the ULS23 challenge \cite{Grauw2024}, referred to as ULS23 DeepLesion.
    \item 697 bone oncological lesions and 120 pancreatic tumors from the Radboudumc hospital, available through the ULS23 dataset \cite{Grauw2024}.
    \item 470 volumes from the multimodal MSD challenge \cite{Antonelli2022}, using only the Lung, Colon, Pancreas datasets.
    \item 700 lung nodules from the LNDb dataset \cite{Pedrosa2021}.
    \item 300 kidney tumors from the KITS23 dataset \cite{Heller2022}.
    \item 832 liver tumors from the LiTS dataset \cite{Bilic2023}, also part of the MSD challenge.
    \item 932 mediastinal and abdominal lymph nodes from the NIH-LN dataset \cite{Roth2015}.
    \item 2236 lung oncological lesions from the LIDC-IDRI dataset \cite{Armato2011}.
\end{itemize}
A random 90\% training set was selected, with a 10\% held-out test dataset, by analogy with the ULS23 challenge. A randomly selected set of 67 tumors $\geq$ 10 mm in size ( $\geq$ 15 mm for lymph nodes) from the ULS23 DeepLesion training set was kept out for validation against radiologists.

\subsection{Segmentation Process}
The model had access to the entire volume and to the visual prompt. The image thresholding was fixed at -500 ; +1000 UH, an unrestricted window akin to bone windowing, which empirically led to the best overall results (data not shown). The model outputs then an initial segmentation mask in 2D. The segmentation masks for the remaining 2D axial slices are then calculated autoregressively, using the mask from the adjacent slice as the prompt for the next slice. This process propagates the segmentation masks from the middle slice, resulting in a 3D segmentation mask.

The ONCOPILOT model is evaluated in three experimental settings using visual prompts that simulate real-life usage:
\begin{itemize}
    \item \textbf{Bounding box}: the model is prompted with a 2D bounding box outlining the lesion from the middle slice of the ground-truth mask, expanded with an offset of 15 pixels. 
    \item \textbf{Point-click}: the model is prompted by a single point, which is determined as the barycenter of the ground-truth mask or the closest point that falls within the segmentation mask. 
    \item \textbf{Point-edit}: to simulate interactions with radiologists, the 3D segmentation mask proposed in point-click mode is refined using up to 4 edit point-clicks chosen as the barycenter of the prediction error that can be negative (i.e., reduce an over-segmented mask) or positive (i.e., expand an under-segmented mask).
\end{itemize}

\begin{figure*}[htbp]
    \makebox[\textwidth][c]{\includegraphics[width=0.9\textwidth]{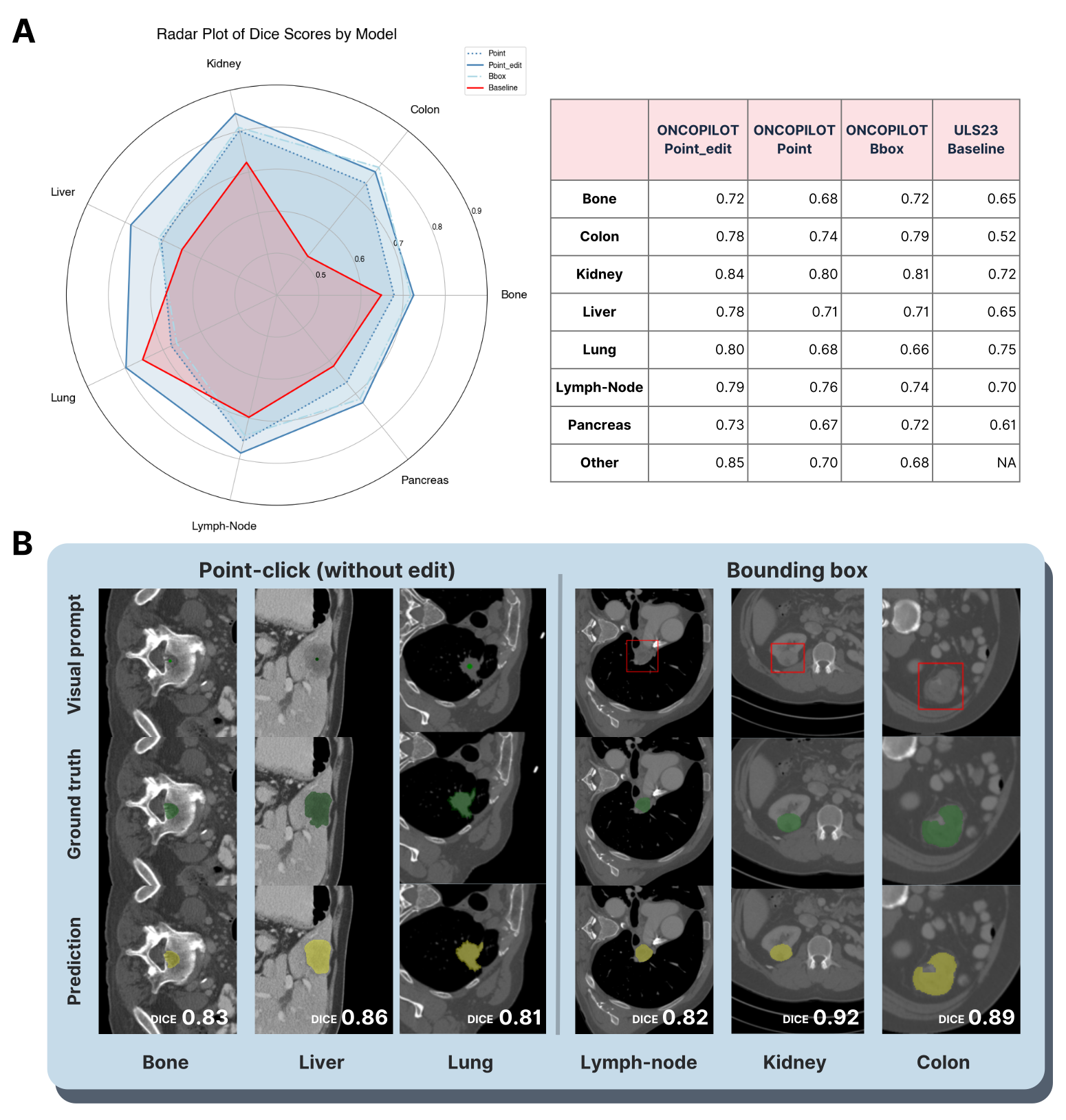}}
    \captionsetup{width=1\textwidth, labelfont=bf}
    \caption{\textbf{ONCOPILOT Performance Against Baseline} (A) Radar plot (top) and table (bottom) displaying segmentation DICE scores across 7 lesion types for 3 different ONCOPILOT models (point, point-edit, bbox) compared to the best-performing baseline from the ULS23 segmentation challenge on the 10\% held-out test set. (B) Examples of successful segmentations from the test set, comparing point mode (left columns) and bbox mode (right columns). The top row shows the visual prompt provided to the model, the middle row displays the ground truth mask for that slice, and the bottom row presents the ONCOPILOT model’s predicted segmentation.}
    \label{fig:fig2}
\end{figure*}

\subsection{Morphology Analysis}
Sphericity index is calculated as the ratio of the surface area of a sphere to the surface area of the ground truth segmentation mask, given equal volumes, with a perfect sphere having a sphericity index of 1 while irregular structures being closer to 0. The following formula was used:
\[
S = \frac{\pi^{1/3} \cdot (6V)^{2/3}}{A}
\]
where \(S\) is the sphericity, \(V\) is the volume of the object, and \(A\) is the surface area of the object.

\subsection{RECIST Measurement}
RECIST measurements from ONCOPILOT were inferred from the segmentation masks in bounding box, point and point-edit modes. The primary measurement evaluated was the long axis of the oncological lesion, with the following amendments for simplicity and consistency: it was applied even to lymph nodes and restricted to the axial plane.

To further understand ONCOPILOT's performance in real oncological evaluations, we compared it against a panel of radiologists for RECIST v1.1 measurements. Using a validation set of 67 tumors from diverse organs kept out from the ULS23 DeepLesion dataset, we compared the long axis in the axial plane inferred from ONCOPILOT predicted segmentation masks to manual annotations made by the radiologist panel composed of three radiologists with a minimum of 18 months of experience. These 67 tumors are selected for their inclusivity as potential target lesions according to the RECIST v1.1 guidelines (i.e., solid lesions with a long axis $\geq$ 10 mm, lymph nodes with a short axis $\geq$ 15 mm) and filtered for their segmentation quality. The measures proposed by ONCOPILOT and the radiologists were compared to those inferred from the ground-truth segmentation masks to extract the measurement error.

Radiologists used our in-house viewer for manual and ONCOPILOT-assisted measurement of the lesion's long axis. They could zoom at will, modify the image's windowing, and navigate the volume freely, without the help of multi-planar reconstruction. The barycenter of the lesion was superimposed on the initial volume to indicate the lesion of interest without biasing the radiologist by showing the ground-truth mask. For ONCOPILOT-assisted measurements, the radiologists used the measures inferred by the model after automatic segmentation of the target lesion by ONCOPILOT using bounding box visual prompts only for simplicity. 

Inter-operator variability was defined as the deviation of each radiologist's measurement from the overall average of all measurements for a given lesion, whether those measurements are aided by the segmentation model or done manually. Essentially, it represents the mean error of each radiologist compared to the collective average. Measurement duration was recorded for each assessment. It is defined as the time from the CT's initial display to getting the object's final measurement.

\vspace{1em}

\begin{figure*}[htbp]
    \centering
    \makebox[\textwidth][c]{\includegraphics[width=0.55\textwidth]{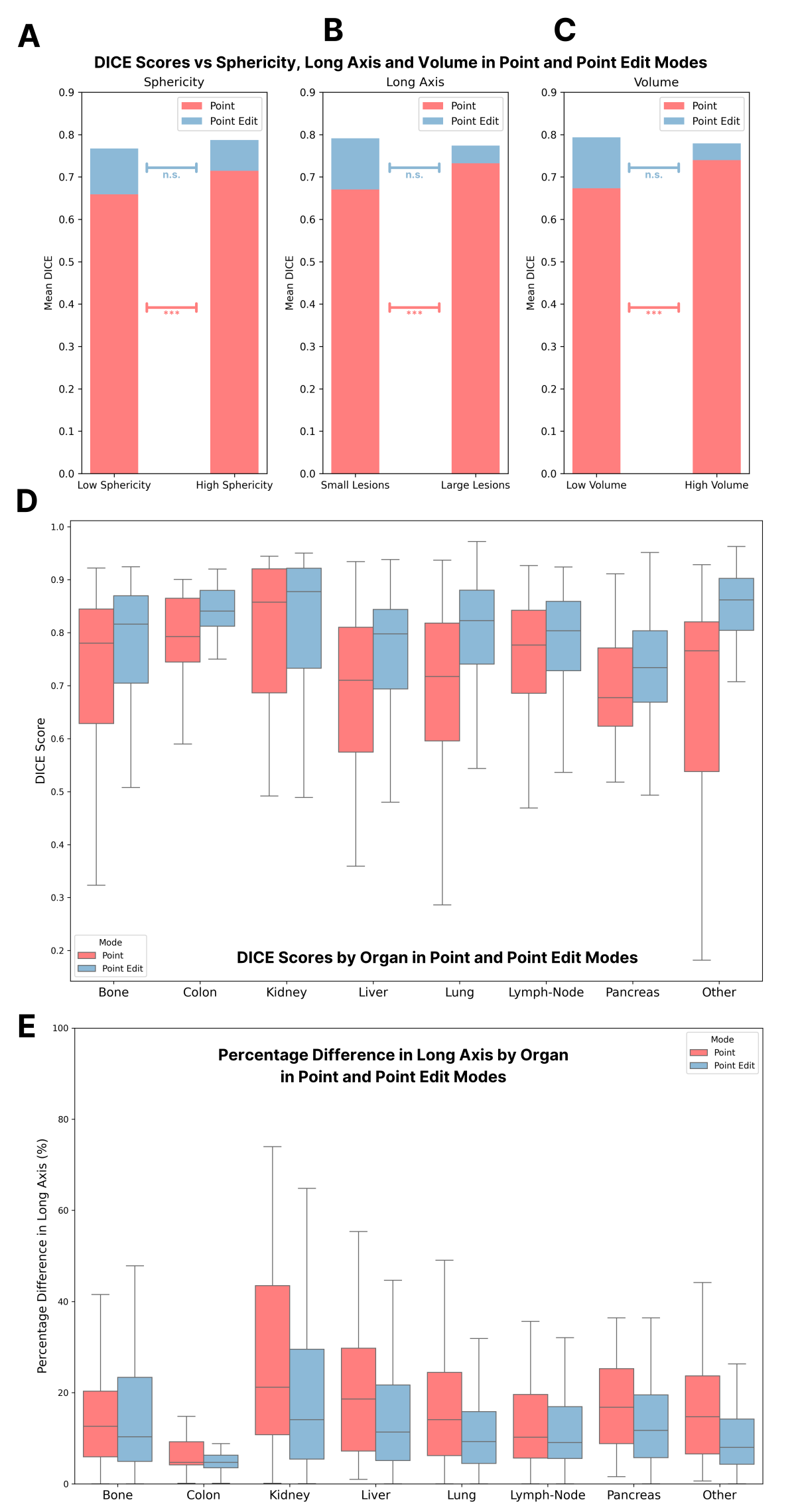}}
    \captionsetup{width=1.1\textwidth, labelfont=bf}
    \caption{\textbf{ONCOPILOT Performance on Different Lesion Types} Bar plot showing the mean DICE scores from ONCOPILOT segmentation masks in point mode (red) and point-edit mode (blue) for: (A) spherical lesions (sphericity $> 0.6$) versus irregular lesions (see Methods for the sphericity formula), (B) large lesions (long axis $> 15$ mm) versus smaller lesions, (C) voluminous lesions (volume $> 1$ mL) versus smaller lesions. (D) Boxplot displaying the distribution of DICE scores produced by ONCOPILOT in point mode (red) and point-edit mode (blue) across various lesion types in the 10\% held-out test set, with median values and interquartile ranges highlighted. (E) Boxplot showing RECIST measurements derived from ONCOPILOT’s predicted masks in point mode (red) and point-edit mode (blue) across different lesion types in the 10\% held-out test set, highlighting median values and interquartile ranges. The long axis is defined as the longest possible line in the axial plane across the predicted 3D mask. ***: p-value $< 0.001$; n.s.: non-significant.}
    \label{fig:fig3}
\end{figure*}

\subsection{ONCOPILOT integration into radiologist’s workflow}

Finally, to determine whether ONCOPILOT could serve as an AI companion, we evaluated its integration and improvement of the oncological evaluation workflow within our in-house environment. We measured inter-reader variability as the mean deviation from the average between radiologists performing RECIST measurements manually and radiologists performing these measures semi-automatically using the segmentation model by drawing a bounding box around the lesion. We also timed the duration required to obtain measurements in each scenario.

\section{Results}

\subsection{Foundation Model}
Our foundation model, ONCOPILOT, was pre-trained on a diverse dataset comprising normal anatomy and oncological lesions, totaling 2,374 CT scans including 104 anatomical structures (e.g., organs, bones) and 4 oncological lesions regardless of histology and malignity (i.e., lung, liver, pancreas and colon tumors) from the MSD dataset (Figure 1A), without distinction regarding their histological type or malignancy.

To become specialized for oncology the model was subsequently fine-tuned on a comprehensive dataset of 6,229 tumors from various organs (e.g., pancreas, bone, liver, kidney, lung, lymph nodes). ONCOPILOT is designed to interactively segment oncological lesions in 3D, utilizing visual prompts such as a bounding box (referred to as bbox) around the lesion of interest or a point-click (referred to as point) inside it (Figure 1B). To simulate the dynamic refinement of the predicted segmentation masks by radiologists we developed an editing mechanism (referred to as point-edit, see Material \& Methods).

\subsection{Segmentation Performance}

ONCOPILOT surpassed the baseline model in all evaluation metrics—point, point-edit, and bbox—across all lesion types, with the exception of lung tumors, where only the point-edit model demonstrated superior performance (Figure 2A, with examples of successful segmentations in Figure 2B).

The test dataset was imbalanced, with over 40\% of the lesions being lung tumors, which biased the overall DICE score in favor of the baseline model (further adressed in the Discussion section). ONCOPILOT achieved mean DICE scores of 0.70 for point mode, 0.70 for bbox mode, and 0.78 for point-edit mode, compared to 0.70 for the baseline. The distribution of lesion sizes by organ and examples of failed segmentations are provided in Supplementary Figures S1A and S1B. Additionally, it is worth mentioning that nnUNet models are often restricted to specific tasks, tends to underperform on complex datasets, and require longer training and inference times while ONCOPILOT offers an all-in-one solution.

\subsection{Morphology Analysis}

The segmentation masks outputted by the model in point mode were influenced by the lesion morphology and size. Indeed, ONCOPILOT exhibited lower DICE scores for lesions with irregular, non-spherical shapes, with a mean DICE of 0.66 for tumors with a sphericity index below 0.6, compared to 0.71 for more spherical tumors in point mode ($p < 0.001$, Figure 3A, Supplementary Figure S2A).

Similarly, smaller lesions yielded lower DICE scores, with a mean of 0.67 for lesions with a long axis $< 15 mm$ versus 0.73 for larger lesions ($p < 0.001$, Figure 3B, Supplementary Figure S2B). This trend persisted when using volume as a metric: lesions under 1 mL had a mean DICE of 0.67, compared to 0.74 for larger lesions ($p < 0.001$, Figure 3C, Supplementary Figure S2C). Crucially, interactive editing mitigated these biases, eliminating significant differences in DICE scores between lesions of varying sphericity, long axis, or volume in point-edit mode. This approach also reduced disparities in DICE between lesion types (Figure 3D). Additionally, when using RECIST measurements for the long axis instead of DICE scores, interactive editing significantly reduced measurement errors, with the median error decreasing from 14.1\% in point mode to 9.6\% in point-edit mode ($p < 0.001$). This level of accuracy is consistent with the reported inter-reader variability among radiologists for single-lesion measurements (\citep{Yoon2016}, Figure 3E).

\subsection{ONCOPILOT Evaluation Against Radiologists}

ONCOPILOT demonstrated radiologist-level performance in point, point-edit, and bbox modes (Figure 4A, 4B). There was no statistically significant difference between the different ONCOPILOT models when evaluated against radiologists, with a median absolute error in long axis measurement of 1.3 mm for radiologists (8.6\% of the median lesion size) versus 1.1 mm for ONCOPILOT in point-edit mode (7.4\%), 1.6 mm in point mode (10.8\%), and 1.5 mm in bbox mode (10.4\%).

\subsection{ONCOPILOT Integration into Radiologist’s Workflow}

ONCOPILOT enhanced the reproducibility and efficiency of radiologist measurements, with an inter-reader deviation of 1.7 mm when assisted by ONCOPILOT versus 2.4 mm manually (Figure 4C, 4D, $p < 0.05$). Additionally, radiologists demonstrated a faster measurement speed using ONCOPILOT, with an average time of 17.2 seconds per measurement compared to 20.6 seconds with manual annotations ($p < 0.05$). Notably, this improvement in speed was achieved without focusing on speed optimization, as it operated on a non-optimized web-based platform (showcased in Supplementary Figure S3A to S3E). Most of the measurement time was spent locating the lesion within the exam, suggesting that ONCOPILOT could be further accelerated with targeted improvements.

\section{Discussion}

In summary, ONCOPILOT demonstrated state-of-the-art performance in tumor segmentation across a diverse set of oncological lesions, achieving radiologist-level accuracy in RECIST 1.1 measurements. The model's flexibility, enabled by interactive visual prompts and refinement capabilities in our in-house viewer, marks a significant advancement in integrating an explainable AI copilot into the imaging workflow while keeping the radiologist in the loop. This approach not only reduces inter-reader variability and measurement time but also offers a more adaptable solution compared to rigid, traditional segmentation models. Nonetheless, further research is needed to establish how these gains in efficiency and precision translate into meaningful improvements in longitudinal oncological evaluation and influence disease status assessment.

ONCOPILOT not only enhances the precision and consistency of RECIST-based oncological assessments but also goes beyond traditional RECIST measurements by enabling volumetric analysis and uncovering previously unexplored radiomic features. Volumetric biomarkers, such as tumor growth rate and total tumor burden, combined with morphology-based markers, offer more comprehensive and accurate indicators of tumor mass and aggressiveness compared to conventional long and short axis measurements. These novel radiomic biomarkers will better accommodate the variability in tumor presentations, providing a more precise characterization of oncological disease.

\begin{figure*}[htbp]
    \centering
    \makebox[\textwidth][c]{\includegraphics[width=1\textwidth]{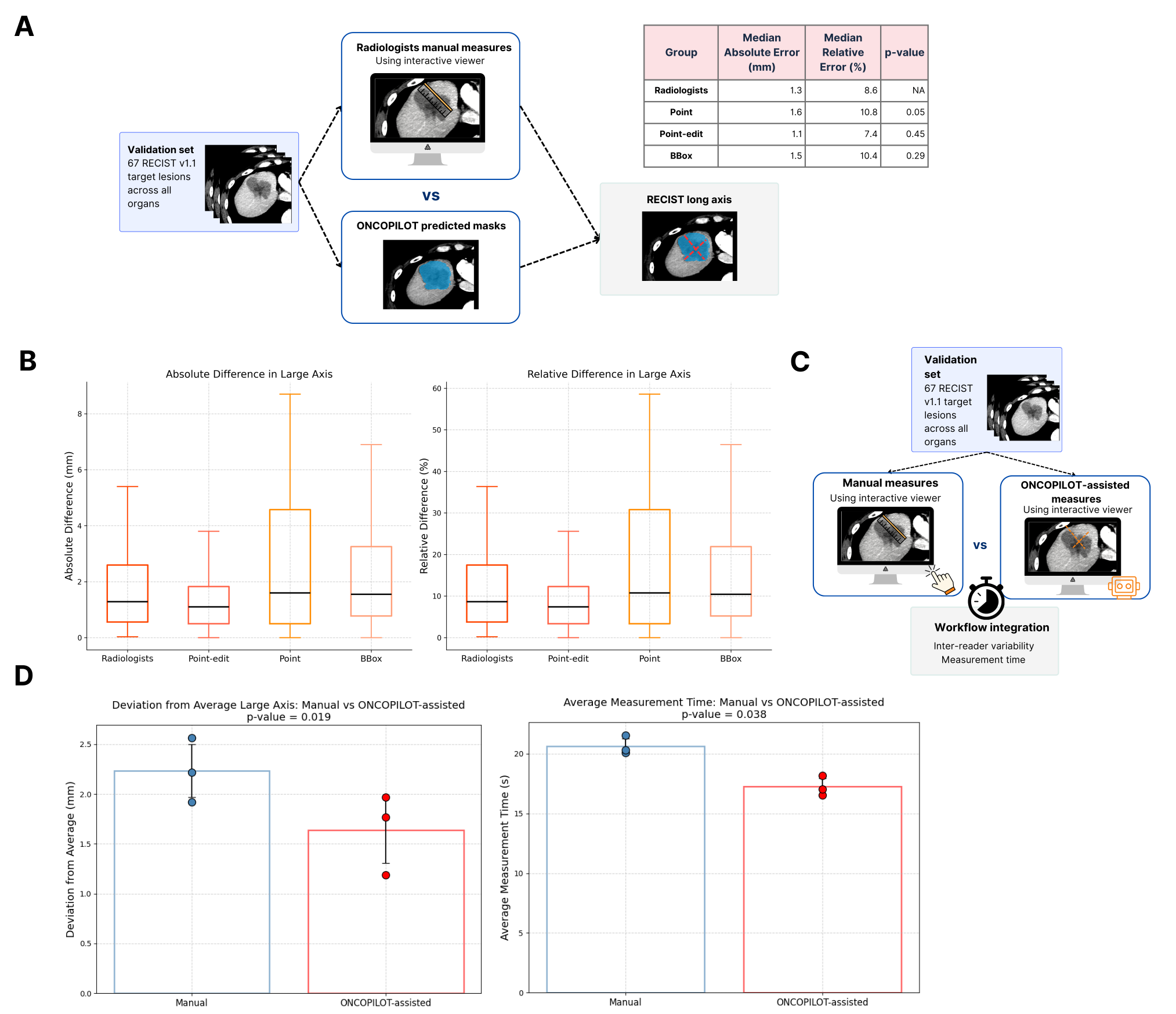}}
    \captionsetup{width=1.1\textwidth, labelfont=bf}
    \caption{\textbf{ONCOPILOT Integration Into Radiologist’s Workflow} (A) Diagram and results comparing ONCOPILOT in point, point-edit, and bbox modes against three radiologists for the long-axis measurement of diverse oncological lesions. Median absolute error (mm) and median relative error (\% of lesion size) are shown. P-values from t-tests compare ONCOPILOT models to radiologists for long-axis measurement error, without statistical significance p $\geq$ 0.05. The long axis is the longest line in the axial plane across the predicted 3D mask. (B) Boxplot (bottom) of ONCOPILOT’s tumors long-axis measurement performance against radiologists. Left: median absolute error (mm) vs. ground truth. Right: median relative error (\% of lesion size). Median and interquartile ranges are shown. (C) Diagram of an experiment evaluating radiologists' inter-operator variability and measurement time while measuring tumors' long-axis using a digital viewer for manual vs. ONCOPILOT-assisted (bbox mode) long-axis assessments. (D) Boxplots show radiologists' inter-operator variability in measurement error (left) and measurement time (right) using manual vs. ONCOPILOT-assisted annotations across diverse tumors, with t-test p-values; n=3.}
    \label{fig:fig4}
\end{figure*}

It is important to note that ONCOPILOT's suboptimal performance on lung tumors appears to be related to lesion size, highlighting a limitation of our model. Lung tumors in the test set were predominantly small nodules of uncertain oncological relevance, with a median size of 9 mm compared to 20 mm for non-lung tumors. This disproportionate representation (more than 40\%) of lung lesions in the test dataset skewed the overall results, disadvantaging our model's performance. Future versions of the model, along with a more balanced test dataset, should address this limitation.

ONCOPILOT leveraged publicly available data \citep{Grauw2024} and model architecture \citep{Grauw2024,Kirillov2023}, demonstrating that foundation models are already capable of delivering impactful results in the biomedical field without significant technical hurdles. Although ONCOPILOT can be considered a small foundation model in terms of training set size, it already showcases the promising potential of this technology, with future iterations expected to be significantly more advanced and effective. These results reinforce our belief that foundation models are a pivotal step toward the next generation of AI-assisted radiology.

\newpage
Through this work, we aim to demonstrate oncological evaluation as the first use case for the native integration of foundation-model-based AI assistants into the radiologist’s workflow, paving the way for improved patient stratification, optimized clinical trial monitoring, more informed treatment decisions, and ultimately enhanced patient care.

\section*{Acknowledgments}
This work was granted access to the HPC resources of IDRIS under the allocation 2024-AD011013489R2 made by GENCI.

\FloatBarrier 

\bibliography{main}
\bibliographystyle{unsrt}

\onecolumn

\newpage

\section*{Supplementary Figures}
\begin{figure*}[h]
    \centering
    \renewcommand\thefigure{S1}  
    
    \begin{minipage}{1\textwidth}
        \centering
        \makebox[\textwidth][c]{\includegraphics[width=1\textwidth]{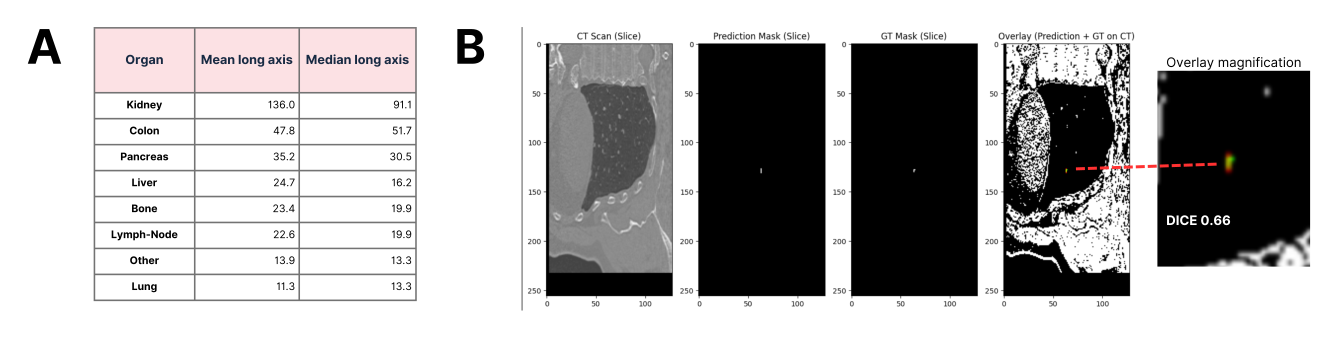}}
        \captionsetup{width=1\textwidth, labelfont=bf}
        \caption{\textbf{ONCOPILOT Long Axis Performance Across Different Organs} (A) Table showing the mean and median long-axis measurements (in mm) for the various organ types in the test set. (B) Example of a suboptimal segmentation by ONCOPILOT on a small lung nodule from the LIDC-IDRI dataset with magnification of the overlay on the rightmost panel, with a DICE of 0.66 in point mode.}
        \label{fig:figs1}
    \end{minipage}
    
    \vspace{1em}   
    \renewcommand\thefigure{S2}  
    
    \begin{minipage}{1\textwidth}
        \centering
        \makebox[\textwidth][c]{\includegraphics[width=1\textwidth]{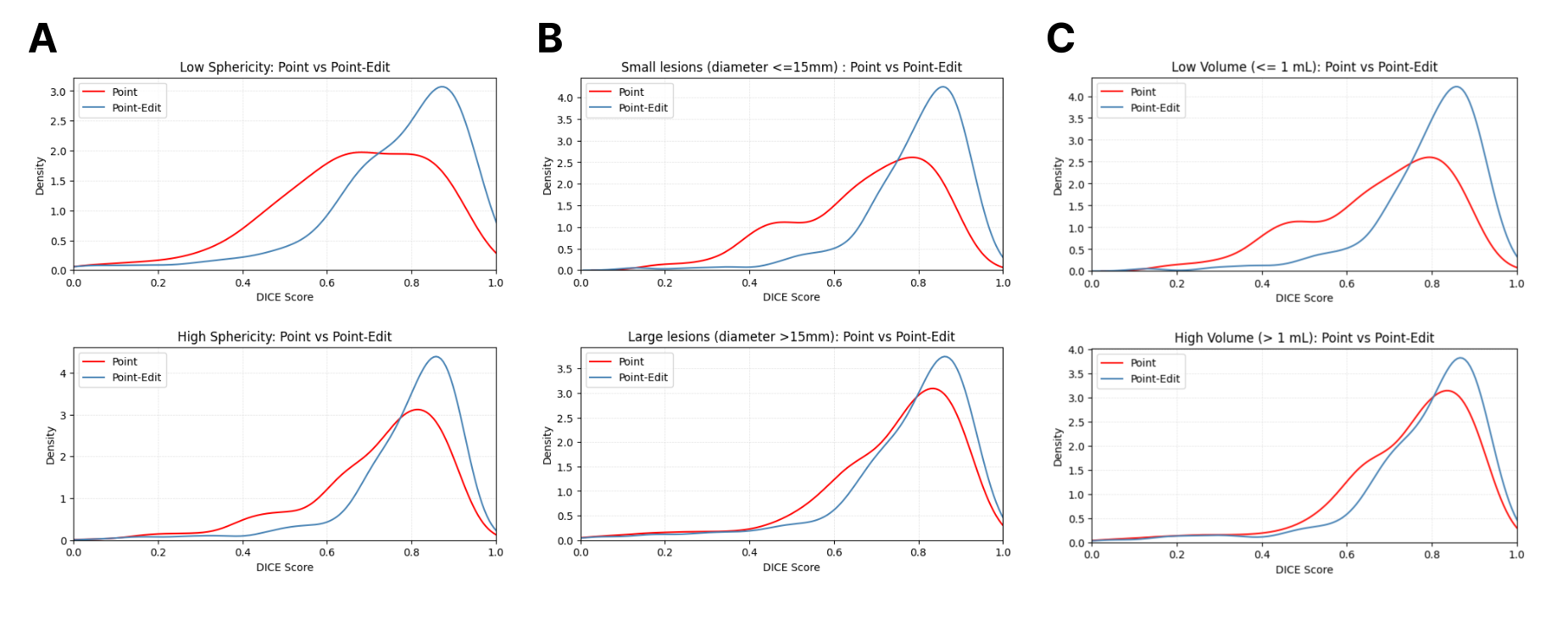}}
        \captionsetup{width=1\textwidth, labelfont=bf}
        \caption{\textbf{ONCOPILOT Morphology Evaluation} (A) Density plot showing the distribution of DICE scores from ONCOPILOT segmentation masks in point mode (red) and point-edit mode (blue) for spherical lesions (sphericity $> 0.6$) versus irregular lesions (see Methods for the sphericity formula). (B) Density plot showing the distribution of DICE scores from ONCOPILOT segmentation masks in point mode (red) and point-edit mode (blue) for large lesions (long axis $> 15$ mm) versus smaller lesions. (C) Density plot showing the distribution of DICE scores from ONCOPILOT segmentation masks in point mode (red) and point-edit mode (blue) for voluminous lesions (volume $> 1$ mL) versus smaller lesions.}
        \label{fig:figs2}
    \end{minipage}

\end{figure*}

\renewcommand\thefigure{S3}  

\begin{figure*}[h]
    \centering
    \makebox[\textwidth][c]{\includegraphics[width=1\textwidth]{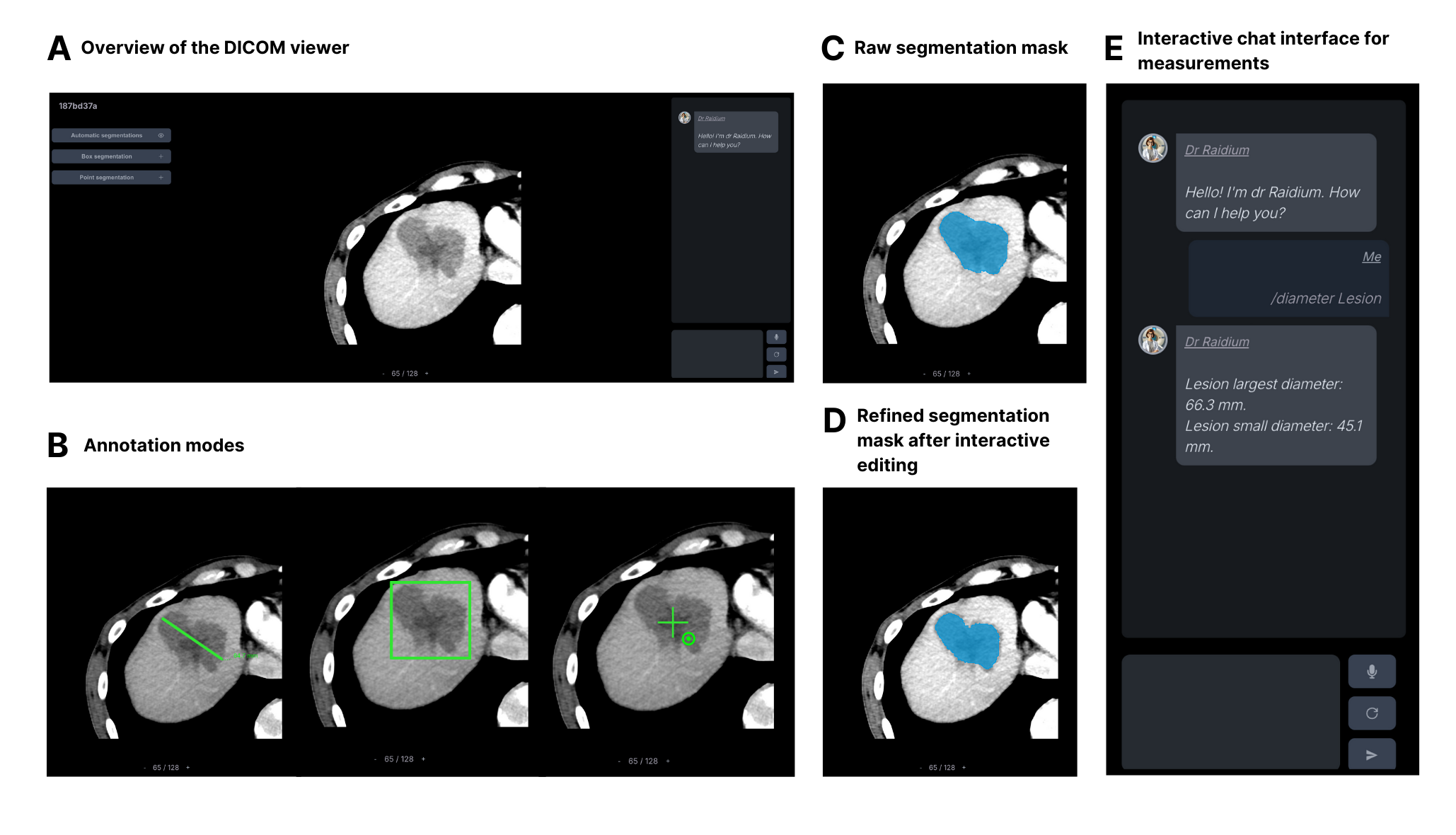}}
    \captionsetup{width=1\textwidth, labelfont=bf}
    \caption{\textbf{ONCOPILOT Integration into a DICOM Viewer} (A) Overview of our in-house viewer featuring ONCOPILOT segmentation options and an interactive chat window for queries. A hypoattenuating liver lesion to be measured is depicted. (B) Depiction of different measurement modes: left panel : manual long-axis measurement; middle panel : bounding-box segmentation; right panel : point-click segmentation. (C) A proposed segmentation mask following bounding-box segmentation, showing slight over-segmentation at the anterior part of the lesion. (D) A proposed segmentation mask after interactive editing with the user to correct over-segmentation. (E) Interactive retrieval of measurements through a chat window.}
    \label{fig:figs3}
\end{figure*}

\newpage

\end{document}